\documentclass[12pt]{article}
\usepackage{amsmath}
\usepackage{graphicx}
\begin{document}
\baselineskip=18 pt
\begin{center}
{\large{\bf A cylindrically symmetric, static anisotropic fluid spacetime and the naked singularity}}
\end{center}

\vspace{.5cm}

\begin{center}
{\bf Faizuddin Ahmed}\footnote{Corresponding author : faizuddinahmed15@gmail.com}\\
{\it Ajmal College of Arts and Science, Dhubri-783324, Assam, India,}\\
{\bf Farook Rahaman}\footnote{rahaman@associates.iucaa.in;farookrahaman@gmail.com}\\
{\it Department of Mathematics, Jadavpur University, Kolkata-700032, West Bengal, India}
\end{center}

\vspace{.5cm}

\begin{abstract}

In this article, a cylindrical symmetry and static solution of the Einstein's field equations, was presented. The spacetime is conformally flat, regular everywhere except on the symmetry axis where it possesses a naked curvature singularity. The matter-energy source anisotropic fluids violates the weak energy condition (WEC), diverge on the symmetry axis. We discuss geodesics motion of free test-particles near to the singularity, and geodesic expansion in the metric to understand the nature of singularity which is naked or covered, and finally the C-energy of the spacetime. 

\end{abstract}

{\it Keywords:} exact solution, cylindrical symmetry, anisotropic fluid, naked singularity

\vspace{0.2cm}

{\it PACS numbers}: 04.20.Jb, 40.20.-q, 04.20.Dw

\vspace{.5cm}

\section{Introduction}

Investigation of the nature of singularities in gravitational collapse solution of the Einstein's field equations in different systems are of particular in general relativity. The study of gravitational collapse in spherically symmetric spacetime has led to many examples of naked singularities ({\it e.g.} \cite{Vaid,Chri}, see also \cite{Jos,Krasin} and references therein). In addition, cylindrical symmetric system have great interest because of one degree of freedom for gravitational waves. Therefore, gravitational collapse in cylindrical symmetric system is the simplest one which contains both collapsing body and gravitational waves. A small samples of gravitational collapse model in this system with naked singularities are \cite{Th,Hay,Apo,Eche,Chi,Nakao,Ghosh,Kurita,Nakao2,Nakao3,Gonc,Zade,Wang,Faiz}. Some other non-spherical gravitational collapse model also possesses a naked singularity ({\it e.g.} \cite{Faiz2,Faiz3,Faiz4}). Long back, Thorne proposed a hoop Conjecture \cite{Thor} concerning the formation of black holes in cylindrical symmetry system. However, the general proof of this Conjecture has not yet been known.

In this article, a cylindrical symmetric and static solution of the field equations, possesses a naked curvature singularity on the symmetry axis, satisfying the strong curvature condition, will be presented. Here we have relaxed the Darmois conditions, and the presented solution is a special case of the very well-known static, cylindrical symmetry and conformally flat solutions \cite{Herr1} (see also, \cite{Herr2}).

\section{Conformally flat nonvacuum spacetime}

Consider the following static and cylindrical symmetric spacetime ($x^0=t, x^1=r, x^2=\phi, x^3=z$) given by
\begin{equation}
ds^2=-\sinh^2 r\,dt^2+\cosh^2 r\,dr^2+\sinh^2 r\,d\phi^2+\sinh^2 r\,dz^2,
\label{1}
\end{equation}
where we have chosen the speed of light $c=1$, the Planck's constant $h=1$, and the gravitational constant $G=1$ such that individual term in the line element has same dimension. The ranges of the coordinates are 
\begin{equation}
-\infty < t < \infty,\quad 0 \leq r < \infty,\quad \phi\in[0,2\,\pi), \quad -\infty < z < \infty.
\label{range}
\end{equation}
The metric has signature $+2$ and its determinant
\begin{equation}
det\;g=-\sinh^{6} r\,\cosh^{2} r,
\label{3}
\end{equation}
degenerates on the symmetry axis at $r=0$. The presented spacetime (\ref{1}) is a nonvacuum solution of the field equations and the non-zero components of the Einstein tensor are
\begin{equation}
G^{0}_{0}=\frac{1}{3}\,G^{1}_{1}=G^{2}_{2}=G^{3}_{3}=\mbox{csch}^{2} r.
\label{4}
\end{equation}
The Ricci scalar (curvature scalar) $R$ of the metric (\ref{1}) is
\begin{equation}
R^{\mu}_{\mu}=R=-6\,\mbox{csch}^{2} r.
\label{ricci}
\end{equation}
And the Kretschmann's curvature scalar is
\begin{equation}
K=R^{\mu\nu\rho\sigma}\,R_{\mu\nu\rho\sigma}=12\,\mbox{csch}^{4} r.
\label{6}
\end{equation}
From above, it is clear that the Kretschmann's curvature scalar (K) diverge on the symmetry axis at $r=0$, vanish rapidly at spatial infinity $r\rightarrow \infty$ along the radial direction. Since the curvature singularity occurs without an event horizon, and the presented solution does not represent a black holes. Therefore, the singularity which is formed due to scalar curvature in a region non-covered by an event horizon is naked.

The presented spacetime (\ref{1}) satisfy the following conditions :

\vspace{0.1cm}

(i) {\it The existence of an axially symmetric axis} \cite{Mars,Mars2,Carot}: The spacetime that has an axially symmetric axis is assured by the condition,
\begin{equation}
\boldsymbol{X}=||\partial_{\phi}||^2=|g(\partial_{\phi},\partial_{\phi})|=|g_{\phi\phi}|=|\sinh^2 r|\rightarrow 0,
\label{cond1}
\end{equation}
as $r\rightarrow 0^{+}$, where we have chosen the radial coordinate such that the axis is located at $r=0$. Here $\partial_{\phi}$ is a spacelike Killing vector fields, the generator of axial symmetry along the cylinder whose orbit is closed. 

(ii) {\it The elementary flatness on the axis} \cite{Mars2,Carot,Carot2,Steph} : This condition implies that the spacetime be locally flat on the axis, which can be expressed as,
\begin{equation}
\frac{(\nabla_{\mu}{\boldsymbol{X}})\,(\nabla^{\mu}{\boldsymbol{X}})}{4\,\boldsymbol{X}}\rightarrow 1,
\label{cond2}
\end{equation}
as $r \rightarrow 0^{+}$. This two conditions ensure the cylindrical symmetry of a spacetime, and $\phi$ denotes the axial coordinate with the hypersurfaces $\phi=0$ and $\phi=2\,\pi$ being identical.

(iii) {\it No closed timelike curves} (CTCs) : We shall consider there is no possibility of occurring closed timelike curves, and simply require that
\begin{equation}
g_{\phi\phi}>0,
\label{cond3}
\end{equation}
hold in all the region of the spacetime considered.

The Einstein field equations (taking cosmological constant $\Lambda=0$) are given by
\begin{equation}
G^{\mu\nu}=\kappa\,T^{\mu\nu},\quad \mu,\nu=0,1,2,3.
\label{field}
\end{equation}
where $\kappa=8\,\pi$ and $T^{\mu\nu}$ is the energy-momentum tensor.

For the presented metric, we consider the following energy-momentum tensor 
\begin{equation}
T_{\mu\nu}=(\rho+p_{\phi})\,U_{\mu}\,U_{\nu}+p_{\phi}\,g_{\mu\nu}+(p_{r}-p_{\phi})\,\eta_{\mu}\,\eta_{\nu}+(p_{z}-p_{\phi})\,\zeta_{\mu}\,\zeta_{\nu},
\label{7}
\end{equation}
where $\rho$ is the energy density, $p_{r}$, $p_{\phi}$ and $p_{z}$ are the principal stresses. Here $U^{\mu}$ is the timelike unit four-velocity vector, $\eta_{\mu}$ and $\zeta_{\mu}$ are the spacelike four vectors along the spatial direction $r$ and $z$, respectively satisfying
\begin{equation}
U^{\mu}\,U_{\mu}=-1,\quad \eta_{\mu}\,\eta^{\mu}=1=\zeta_{\mu}\,\zeta^{\mu},\quad \eta_{\mu}\,U^{\mu}=\zeta_{\mu}\,U^{\mu}=\eta_{\mu}\,\zeta^{\mu}=0.
\label{unit}
\end{equation}
For static fluid distributions, from the spacetime (\ref{1}) we have
\begin{equation}
U_{\mu}=-\sinh r\,\delta^{0}_{\mu},\quad \eta_{\mu}=\cosh r\,\delta^{1}_{\mu},\quad \zeta_{\mu}=\sinh r\,\delta^{3}_{\mu}.
\label{10}
\end{equation}

Therefore, the non-zero components of the energy-momentum tensor from (\ref{7}) using (\ref{10}) are
\begin{equation}
T^{0}_{0}=-\rho,\quad T^{1}_{1}=p_{r},\quad T^{2}_{2}=p_{\phi},\quad T^{3}_{3}=p_{z},
\label{11}
\end{equation}
and the trace of the energy-momentum tensor is
\begin{equation}
T^{\mu}_{\,\mu}=T=-\rho+p_{r}+p_{\phi}+p_{z}.
\label{12}
\end{equation}
Equating the non-zero components of the field equations (\ref{field}) using (\ref{4}) and (\ref{11}), one will get
\begin{equation}
-\kappa\,\rho=\mbox{csch}^{2} r,\quad \kappa\,p_{r}=3\,\mbox{csch}^{2} r,\quad \kappa\,p_{\phi}=\kappa\,p_{z}=\mbox{csch}^{2} r,
\label{13}
\end{equation}
where $\kappa=8\,\pi$.

From above, it is clear that the energy-density of anisotropic fluids ($p_{\phi}=p_{z}=p_{t}$ tangential stress) violate the weak energy condition (WEC) \cite{Hawking}. In addition, the physical parameters $\rho$, $p_{r}$, $p_{t}$ are singular (diverge) on the symmetry axis at $r=0$, and vanish rapidly at spatial infinity $r\rightarrow \infty$.

The presented cylindrically symmetric spacetime (\ref{1}) is conformally flat since the Weyl tensor vanishes, {\it i.e.,} $C_{\mu\nu\rho\sigma}=0$ and static. The general integration of the conformally flat condition, for static and cylindrically symmetric case with anisotropic fluids was obtained by L. Herrera {\it et. al.} \cite{Herr1}. The authors there shown that any conformally flat and cylindrically symmetric static source cannot be matched through Darmois conditions to the Levi-Civita spacetime, satisfying the regularity conditions. Furthermore, all static, cylindrical symmetry solutions (conformally flat or not) for anisotropic fluids have been found in \cite{Herr2}. In the present work, we have relaxed the Darmois conditions and the solution is a special case of the solutions \cite{Herr1} (section 4), and also a member of the general solutions exhibited in \cite{Herr2} (section 9.2).

We considered static fluid distributions, there are only three kinematic variables, namely, the {\it expansion} $\Theta$, the {\it acceleration} vector $\dot{U}^{\mu}$, and the {\it shear tensor} $\sigma_{\mu\nu}$ associated with the fluid four-velocity vector which are defined by
\begin{eqnarray}
\Theta&=&U^{\mu}_{\,\,;\,\mu},\nonumber\\
a^{\mu}&=&\dot{U}^{\mu}=U^{\mu\,;\,\nu}\,U_{\nu},\nonumber\\
\sigma_{\mu\nu}&=&U_{(\mu\,;\,\nu)}+\dot{U}_{(\mu}\,U_{\nu)}-\frac{1}{3}\,\Theta\,h_{\mu\nu},
\label{15}
\end{eqnarray}
where $h_{\mu\nu}=g_{\mu\nu}+U_{\mu}\,U_{\nu}$ is the projection tensor. For the metric (\ref{1}), these parameter have the following expressions :
\begin{equation}
\Theta=0=\sigma_{\mu\nu},\quad a^{\mu}=\mbox{csch} r\,\mbox{sech} r\,\delta^{\mu}_{r}.
\label{16}
\end{equation}
The magnitude of the acceleration four vector defined by $a^2=a^{\mu}\,a_{\mu}$ is $a=\mbox{csch} r$, which vanish at spatial infinity $r\rightarrow \infty$. The presented cylindrical symmetry, static solution therefore have acceleration, non-expanding, and shear-free.   

Next, we discuss the geodesic motion of free test-particles, nature of singularities and its strength, and the C-energy of the presented spacetime.

\vspace{0.1cm}
\begin{center}
{\bf (A) Geodesics in the neighborhood of the singularity}
\end{center}
\vspace{0.1cm}

The Lagrangian for the metric (\ref{1}) is given by
\begin{eqnarray}
\nonumber
\pounds&=&\frac{1}{2}\,g_{\mu\nu}\,\dot{x}^{\mu}\,\dot{x}^{\nu}\quad,\\
&=& \frac{1}{2}\,\left[-g_{tt}\,\dot{t}^2+g_{rr}\,\dot{r}^2+g_{\phi\phi}\,\dot{\phi}^2+g_{zz}\,\dot{z}^2\right ],
\label{17}
\end{eqnarray}
where dot stands derivative w. r. t. an affine parameter. From (\ref{1}) and (\ref{17}), it is clear that there are exist three constants of motion corresponding to the cyclic coordinates $t$, $\phi$ and $z$. These constants of motion are
\begin{equation}
p_{t}=-E=-g_{tt}\,\dot{t},\quad p_{\phi}=g_{\phi\phi}\,\dot{\phi}=L,\quad p_{z}=g_{zz}\,\dot{z}.
\label{18}
\end{equation}
The normalization condition is
\begin{equation}
g_{\mu\nu}\,\dot{x}^{\mu}\,\dot{x}^{\nu}=\delta,
\label{19}
\end{equation}
where $\delta=0$ for null geodesics and $\delta=-1$ for time-like.

For null geodesics in the $z=const$-planes, we have from the normalization condition
\begin{eqnarray}
-g_{tt}\,\dot{t}^2+g_{rr}\,\dot{r}^2+g_{\phi\phi}\,\dot{\phi}^2=0\nonumber\\
\Rightarrow \dot{r}^2=\frac{(E^2-L^2)}{g_{rr}\,g_{\phi\phi}}\nonumber\\
\Rightarrow \sinh r\,\cosh r\,\dot{r}=\sqrt{E^2-L^2}.
\label{20}
\end{eqnarray}
The solution of above equation is
\begin{equation}
r(\lambda)=\sinh^{-1} [\{2\,\sqrt{E^2-L^2}\,\lambda\}^{1/2}], \quad E>L,
\label{22}
\end{equation}
where the constant of integration is taken zero.

\begin{figure}[htbp]
    \centering
        \includegraphics[scale=.35]{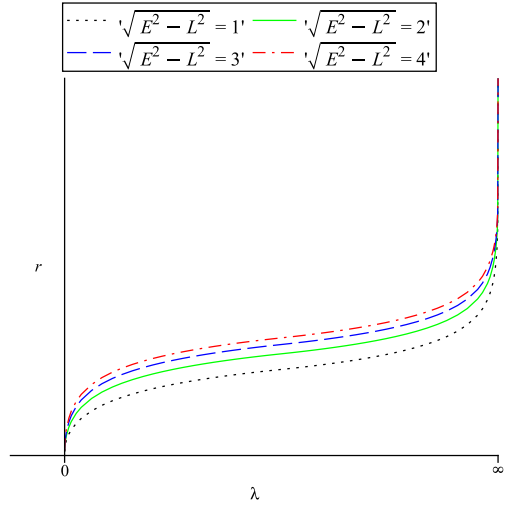}
       \caption{Null geodesics for different values of parameter.  }
  \end{figure}

For the radial null geodesics $\dot{z}=0=\dot{\phi}$, we have from (\ref{19})
\begin{equation}
-g_{tt}\,\dot{t}^2+g_{rr}\,\dot{r}^2=0\Rightarrow \dot{r}^2=\frac{E^2}{g_{rr}\,g_{\phi\phi}}.
\label{21}
\end{equation}
The solution of eqn. (\ref{21}) is
\begin{equation}
r(\lambda)=\sinh^{-1} [\sqrt{2\,E\,\lambda}].
\label{23}
\end{equation}
Therefore, the first derivative of $t$ with its solution from eqn. (\ref{18}) for the null geodesics is
\begin{eqnarray}
\dot{t}=\frac{E}{g_{\phi\phi}}&=&\frac{E}{\sinh^{2} r}=\frac{E}{2\,\sqrt{E^2-L^2}\,\lambda},\nonumber\\
t(\lambda)&=&\frac{E}{2\,\sqrt{E^2-L^2}}\,\mbox{ln} \lambda.
\label{24}
\end{eqnarray}
Therefore, from eqn. (\ref{24}) it is clear that time coordinate $t$ for the null geodesics is incomplete {\it i.e.} the null geodesics coming from infinity hit the singularity $r=0$ (see fig 1).

\vspace{0.1cm}
\begin{center}
{\bf (B) Geodesic expansion in the metric}
\end{center}
\vspace{0.1cm}

Consider a congruence of radial null geodesics in the presented space-time, having the tangent vectors $K^{\mu}(r)=(K^{t},K^{r},0,0)$, where $K^{t}=\frac{dt}{d\lambda}$, and $K^{r}=\frac{dr}{d\lambda}$. The geodesic expansion is given by \cite{SINGH}
\begin{equation}
\boldsymbol{\Theta}=K^{\mu}_{\,\,;\,\mu}=\frac{1}{\sqrt{-g}}\,\frac{\partial}{\partial\,x^{i}}\,(\sqrt{-g}\,K^{i}).
\label{25}
\end{equation}
For the presented metric, one will get from above 
\begin{equation}
\boldsymbol{\Theta}=\frac{\partial K^{r}}{\partial r}+\frac{K^{r}}{\sqrt{-g}}\,(\sqrt{-g})^{'}.
\label{26}
\end{equation}
where prime denotes derivative w. r. t. $r$. We proceed by noting that
\begin{eqnarray}
\frac{dK^{r}}{d\lambda}&=&\frac{\partial K^{r}}{\partial r}\,\frac{\partial r}{\partial \lambda},\nonumber\\
\frac{dK^{t}}{d\lambda}&=&\frac{\partial K^{t}}{\partial r}\,\frac{\partial r}{\partial \lambda}.
\label{27}
\end{eqnarray}
Therefore, we get the final expression of the geodesic expansion (\ref{26})
\begin{eqnarray}
\boldsymbol{\Theta}(r)&=&\frac{1}{K^{r}}\,\frac{dK^{r}}{d\lambda}+\frac{K^{r}}{\sqrt{-g}}\,(\sqrt{-g})^{'}\nonumber\\
&=&\frac{1}{K^{r}}\,[-\Gamma^{r}_{rr}\,(K^{r})^2-\Gamma^{r}_{tt}\,(K^{t})^2]+\frac{K^{r}}{\sqrt{-g}}\,(\sqrt{-g})^{'}\nonumber\\
&=&-\Gamma^{r}_{rr}\,K^{r}-\frac{(K^{t})^2}{K^{r}}\,\Gamma^{r}_{tt}+\frac{K^{r}}{\sqrt{-g}}\,(\sqrt{-g})^{'},\nonumber\\
&=&\frac{2\,E}{\sinh^{2} r}>0,
\label{28}
\end{eqnarray}
which is positive, a fact which indicates the nature of singularity that is formed due to scalar curvature is naked and would  be observable for far away observers.

According to Clarke {\it et.al.} \cite{Clarke2}, a sufficient condition for the naked singularity to be strong in the sense of Tipler \cite{Tipler,Tipler2} is that
\begin{equation}
\lim_{\lambda\rightarrow 0} \lambda^2\,R_{\mu\nu}\,\frac{dx^{\mu}}{d\lambda}\,\frac{dx^{\nu}}{d\lambda}\neq 0(>0).
\label{29}
\end{equation}
Here  $\frac{dx^{\mu}}{d\lambda}$ is defined as the tangent vector fields to the radial null geodesics and $R_{\mu\nu}$ is the Ricci tensor. The weaker condition (known as the {\it limiting focusing condition} \cite{Kro})  is defined by
\begin{equation}
\lim_{\lambda\rightarrow 0} \lambda\,R_{\mu\nu}\,\frac{dx^{\mu}}{d\lambda}\,\frac{dx^{\nu}}{d\lambda}\neq 0.
\label{30}
\end{equation}

For the presented metric (\ref{1}), we have from condition (\ref{29})
\begin{eqnarray}
&&\lim_{\lambda\rightarrow 0} \lambda^2\,[R_{tt}\,(\frac{dt}{d\lambda})^2+R_{rr}\,(\frac{dr}{d\lambda})^2+R_{\phi\phi}\,(\frac{d\phi}{d\lambda})^2+R_{zz}\,(\frac{dz}{d\lambda})^2]\nonumber\\
&=&\lim_{\lambda\rightarrow 0} 2\,\lambda^2\,(\frac{dt}{d\lambda})^2\nonumber\\
&=&\lim_{\lambda\rightarrow 0} \frac{2\,E^2\,\lambda^2}{\sinh^{4} r}\nonumber\\
&=&\lim_{\lambda\rightarrow 0} (\frac{E^2\,\lambda^2}{2\,E^2\,\lambda^2})\nonumber\\
&>&0,
\end{eqnarray}
where $R_{rr}=0$, $\dot{\phi}=0=\dot{z}$. Similarly, if one can calculate the {\it limiting focusing condition} (\ref{30}), then it becomes zero. Thus the naked singularity (NS) which is formed due to scalar curvature on the axis $r=0$ satisfy the {\it strong curvature condition} only. Therefore, the analytical extension of the space-time through the singularity is not possible.

\vspace{0.1cm}
\begin{center}
{\bf (C) The C-energy of the spacetime}
\end{center}
\vspace{0.1cm}

A cylindrical symmetric spacetime is defined locally by the existence of two commuting, spacelike Killing vectors, where the orbits of one Killing vector is closed and the other one is open. For such a spacetime, there exist coordinates $(\phi, z)$ such that the Killing vectors are $(\xi_{(\phi)}, \xi_{(z)})=(\partial_{\phi}, \partial_{z})$. The metric tensor $g_{\mu\nu}$ therefore is independent of $\phi$ and $z$. The norm of these Killing vectors are invariant, namely, the circumferential radius
\begin{equation}
{\bf r}=\sqrt{|\xi_{(\phi)\mu}\,\xi_{(\phi)\nu}\,g^{\mu\nu}|},
\label{35}
\end{equation}
and the specific length
\begin{equation}
l=\sqrt{|\xi_{(z)\mu}\,\xi_{(z)\nu}\,g^{\mu\nu}|}.
\label{36}
\end{equation}
The gravitational energy per unit specific length in a cylindrical symmetric system (also known as C-energy) as defined by Thorne \cite{Th} is
\begin{equation}
{\bf U}=\frac{1}{8}\,\left(1-\frac{1}{4\,\pi^{2}}\,l^{-2}\,\nabla^{\mu} {\bf A}\,\nabla_{\mu} {\bf A}\right),
\label{37}
\end{equation}
where ${\bf A}={\bf r}\,l$ is the areal radius, and ${\bf U}$ is the C-energy scalar.

For the presented metric, ${\bf r}=\sinh r$ and $l=\sinh r$ so that area of the two-dimensional cylindrical surface ${\bf A}=\sinh^{2} r$. Hence the C-energy scalar ${\bf U}$ is
\begin{equation}
{\bf U}=\frac{1}{8}\,(1-\frac{1}{\pi^{2}})<\frac{1}{8}.
\label{38}
\end{equation}
The C-energy is defined in terms of the C-energy flux vector $P^{\mu}$ which satisfies the conservation law $\nabla_{\mu}\,P^{\mu}=0$. The C-energy flux vector $P^{\mu}$ is defined by
\begin{equation}
P^{\mu}=2\,\pi\,\epsilon^{\mu\nu\rho\sigma}\,{\bf U}_{,\nu}\,\xi_{(z)\rho}\,\xi_{(\phi)\sigma},
\label{flux}
\end{equation}
where $\epsilon^{\mu\nu\rho\sigma}$ is the Levi-Civita skew tensor, and we find the C-energy flux vector $P^{\mu}=0$. The C-energy density measured by an observer whose worldline has a unit tangent $U^{\mu}$ is
\begin{equation}
{\bf E}=-P^{\mu}\,U_{\mu}=0,
\label{flux2}
\end{equation}
and the C-energy flux which the observer sees flowing in a direction $n^{\mu}$ (tangent vector to the singularity) orthogonal to his/her worldline is
\begin{equation}
F=P^{\mu}\,n_{\mu}=0.
\label{flux3}
\end{equation}

\section{Conclusions}

A cylindrical symmetric solution of the Einstein's field equations possesses a naked curvature singularity on the symmetry axis, was presented. The spacetime is conformally flat, static and the matter-energy content anisotropic fluids violate the weak energy condition (WEC). Furthermore, the different physical parameters, the energy density, and pressures are singular on the symmetry axis and vanish rapidly at spatial infinity along the radial direction. All cylindrical symmetric, and conformally flat static solutions, satisfying the regularity conditions are well-known. In this work, we relaxed the Darmois conditions, and the presented solution is a special case of the known solutions \cite{Herr1,Herr2}. Finally, we discussed geodesic expansion in the metric to understand the nature of singularity, and found that the curvature singularity is naked which satisfies the strong curvature condition. Finally, the C-energy of the cylindrical spacetime and the C-energy flux vector was discussed.

\subsection*{Acknowledgments}
The authors would like to thank the anonymous referee for their valuable comments, recommendations and brought out our attention to the important references. We also very much thankful to Prof. Luis Herrera, Central University of Venezuela for remarkable discussion and valuable suggestions. FR would like to thank the authorities of the Inter-University Centre for Astronomy and Astrophysics, Pune, India for providing the research facilities. FR is also thankful to DST-SERB for financial support.

The authors declare that there are no competing interests regarding publication of this paper.

\end{document}